# EQUILIBRIUM CONSTANT AS SOLUTION TO THE OPEN CHEMICAL SYSTEMS

B. Zilbergleyt[I]

The products to reactants ratio in isolated system with only one chemical reaction at true thermodynamic equilibrium (TdE) is equal to equilibrium constant. What happens when the system becomes open and achieves thermodynamic equilibrium with its environment? Answer to this question was recently offered by the Discrete Thermodynamics of Chemical Equilibria (DTD) in form of the logarithmic logistic map [1,2,3]. This paper is intended to show that the map can be transformed into a simple and clear formula to calculate compositions of the open chemical systems at equilibria.

The DTD considers open system a part of the environment and its state as dichotomial subsystem – environment equilibrium. Chemical transformations in the subsystem are driven by the internal and the external thermodynamic forces, the second represents the environment impact on the subsystem. Discrete thermodynamics brings forward their balance as a new principle of chemical equilibrium in open systems. At p,T=const this approach leads to the condition of equilibrium in j-system as a logarithmic logistic map

$$(1) \qquad \ln[\Pi_j(\eta_j,0)/\Pi_j(\eta_j,\delta_j)] - \tau_j\delta_j(1-\delta_j^\pi) = 0,$$

where $\delta_j$ is the subsystem shift from TdE, positive or negative depending upon direction of the move – $\delta_j>0$ to the reactants or $\delta_j<0$ to the products. Factor $\tau_j$ defines the shift "growth" rate. The function under the logarithm is a ratio between traditional molar fraction products at zero shift (isolated system at TdE, $\Pi_j(\eta_{kj},0)$), and at non-zero shift (open system out of TdE, $\Pi_j(\eta_{kj},\delta_j)$). Parameter $\eta_j$ is well known from the law of stoichiometry; in our definition, it is the amount of reaction participant, transformed *ab initio* to TdE, per its stoichiometric unit. Finally, $\pi$ is the system complexity parameter.

Map (1) demands equality of the *full change of Gibbs' free energy* to zero. Its first term is reduced by RT classical Gibbs' free energy change, that guards the classical chemical thermodynamics "family" values. The second is a non-classical parabolic term reflecting the open system interactions with its environment. This term causes a rich variety of "far-from-equilibrium" behaviors, driving the chemical system towards bifurcations and chaos. Map (1) is the *map of states* of chemical systems. As the $\eta_j$ value varies, it opens up as a set of the maps that form the chemical system *domain of states*, extending from true thermodynamic equilibrium to true chaos. Solutions to map (1) are typical pitch-fork bifurcation diagrams in the $\delta_j$–$\tau_j$ coordinates, stemming out of abscissa at different points. Unlike the classical theory, in discrete thermodynamics TdE takes an area on abscissa $\tau_j$ from zero to the TdE limit point, whose extent depends upon the relevant $\eta_j$ value [4]. At $\delta_j=0$ the map turns to classical Gibbs' free energy change for isolated state, and TdE area contracts to a dot.

Now let's transform map (1) first taking antilogs

$$(2) \qquad \Pi_j(\eta_j,0) = \Pi_j(\eta_j,\delta_j)\exp[\tau_j\delta_j(1-\delta_j^\pi)].$$

One can clearly see the operational meaning of this expression: it maps population of the j-system at TdE ($\delta_j=0$) into the same system population at open chemical equilibrium ($\delta_j \neq 0$). Hence, *all states of the chemical system can be deduced from its isolated state at the same values of thermodynamic parameters*. Now, recall that equilibrium constant equals to the mole fraction product $\Pi_j(\eta_j,0)$ at TdE by definition

$$(3) \qquad K_j = \Pi_j(\eta_j,0).$$

Denoting $\mu_j = \tau_j\delta_j(1-\delta_j^\pi)$ for the sake of shortness, after replacement of $\Pi_j(\eta_j,0)$ in (2) with $K_j$ we arrive at the general rule for equilibrium states in chemical systems

$$(4) \qquad \mathbf{K_j = \Pi_j(\eta_j,\delta_j)\exp(\mu_j)}.$$

---

[I] System Dynamics Research Foundation, Chicago, USA, sdrf@ameritech.net

At $\delta_j=0$ we get a trivial solution; otherwise we need to know the $\delta_j$–$\tau_j$ relationship that can be tabulated by programmed iterations. Formula (4) suits equally well the thermodynamic branch, along which the above relationship is unilateral, and bifurcations area with multiple roots. That allows us to find out composition of the chemical system at any $\delta_j$ within the scope $\delta_{jmin}<\delta_j<1$, $\delta_{jmin}<0$ corresponds to the system state beyond TdE with at least one totally consumed reactant. Real chemical systems comprise multiple subsystems; given the resources are limited, joint solution to the set of maps (4), each relevant to a specific subsystem, gives equilibrium composition for each of them. This is a novel approach to thermodynamic simulation of complex chemical systems [4].

Map (4) links traditional constant of equilibrium with the open chemical systems, meaning a fundamental break through in the open systems thermodynamics and leading to formerly unknown opportunities in the analysis of real chemical objects.

Detailed derivation, the extended results and their discussion can be found in the references.